\documentclass[
twocolumn,
preprintnumbers,
nofootinbib,
prd,aps
]{revtex4}

\usepackage{epsf}
\usepackage{latexsym}
\usepackage{amssymb}
\usepackage{epsfig}
\usepackage{amsmath}
\usepackage{lscape}

\begin{document}


\newcommand{\rem}[1]{$\spadesuit${\bf #1}$\spadesuit$}
\newcommand{\GeV}{{\text{GeV}}}
\newcommand{\eff}{{\text{eff}}}
\newcommand{\SM}{{\text{SM}}}
\newcommand{\BSM}{{\text{BSM}}}

\renewcommand{\topfraction}{0.8}

\preprint{UT-HET-109,~KU-PH-019}

\title{
Diphoton excess at 750 GeV in an extended scalar sector }

\author{
Shinya Kanemura$^1$, Naoki Machida$^1$, Shinya Odori$^1$, and Tetsuo Shindou$^2$
}
\affiliation{
${}^1$Department of Physics, University of Toyama,
Toyama 930-8555, Japan, \\
${}^2$Division of Liberal-Arts, Kogakuin University,
1-24-2 Nishi-Shinjuku, Tokyo, 163-8677, Japan
}


\begin{abstract}
We discuss an extended scalar model which explains 
the recent results of diphoton excess at 750 GeV at LHC Run II experiments.      
An additional singlet scalar boson with the mass of 750 GeV, which couples to top quarks via a dimension five 
operator, is produced via gluon fusion and decays into two photons via loop 
contributions of a number of (multiply) charged scalar bosons. 
Origin of such a dimension five operator would be, for example, in the context of composite Higgs models. 
The excess can be explained without contradicting the data from LHC Run I and also 
theoretical consistencies such as perturbative unitarity and charge non-breaking.
\end{abstract}

\maketitle

\renewcommand{\thefootnote}{\#\arabic{footnote}}


\section{Introduction}

Since the discovery of the Higgs boson at LHC Run I \cite{higgs}, 
the main target of high energy collider experiments has turned to 
detect new direct evidence of physics beyond the standard model (SM).   
Recently both ATLAS Collaboration and CMS Collaboration reported a new excess in the diphoton data 
at 750 GeV with the width about 45 GeV at LHC Run II~\cite{ATLAS2, CMS2}, 
which might be the resonance of a new particle beyond the SM.  
Many physicists have been trying to understand the new excess based on various ideas, 
and quite a few papers have already been submitted until now for a short time\cite{ref0,ref1}. 
In a large number of the proposed models, 
vector-like fermions are introduced to enhance 
the diphoton decay of new resonance.
Alternatively, there are models in which 
new charged scalar fields in the extended scalar sector 
significantly contribute to the diphoton decay~\cite{ref1}.

In this paper, we would like to discuss a possibility that an extended Higgs sector would explain this phenomenon 
in a relatively simple way.  Although the Higgs boson was found, the shape of the Higgs sector 
remains unknown, and there are many possibilities for extended scalar sectors. 
Such extensions of the Higgs sector are often motivated to understand phenomena which cannot be explained 
in the SM, such as radiative neutrino mass generation mechanisms, sources of a scalar dark matter and the cause of strongly 
first order phase transition and CP violation required for electroweak baryogenesis. 
In addition, new paradigms beyond the SM also require a specific Higgs sector in each model. 

We here introduce a simple extension of the SM with an additional real singlet scalar field $S$ 
and several (multiply) charged scalar bosons. 
We can assume that the singlet does not have a vacuum expectation value (VEV). 
The singlet couples to top quarks  ($S \overline{t}_L t_R$), whose coupling  
originally comes form a dimension five operator $S t_L \tilde{\Phi} t_R$, 
and also couples to charged scalars via trilinear scalar couplings with dimensionful parameters.  
At LHC, the singlet field $S$ can then be produced via the gluon fusion process. The produced $S$ fields 
mainly decay into $t \overline{t}$ but some do into diphoton. 
The observed data\cite{ATLAS2, CMS2} suggest that the signal at $M_{\gamma\gamma} \simeq 750$ GeV should satisfy   
\begin{align}
{\rm ATLAS}~:~
& \sigma(pp \to S X \to \gamma \gamma X)
  \simeq 5 \pm 4~{\rm fb}~(95\% {\rm CL})~,\nonumber \\
{\rm CMS}  ~:~
& \sigma(pp \to S X \to \gamma \gamma X)
\simeq 9 \pm 7~{\rm fb}~(95\% {\rm CL})~.
\end{align}
We show that by our simple setup 
the observed signal cross section and the observed total  
width $\Gamma_S \simeq 45~{\rm GeV}$~\cite{ATLAS2} can be explained 
without contradicting the data from LHC Run I~\cite{ATLAS1, CMS1} and also 
constraints from theoretical consistencies such as perturbative unitarity~\cite{unitarity} 
and charge non-breaking vacuum~\cite{cb}.

\section{Model}

We consider the following effective Lagrangian for interactions of 
the singlet field $S$ as 
\begin{align}
 {\cal L_{\rm eff}} &= - y \frac{S}{\Lambda} 
\left( 
\overline{Q}_L \tilde{\Phi} q_R
\right)   
-  \sum_{q} \sum_{i,j}^{n_q} \mu_q^{ij}~S \phi_i^{+q} \phi^{-q}_j,  \label{eq:Leff}
\end{align}
where $\Phi$ is the Higgs doublet field, $\phi_i^{\pm q}$ are scalar bosons with the electric charge of $\pm q$, and 
$n_q$ is the number of $q$-charged scalar bosons. 
We can assume that the singlet $S$ does not have a VEV while the Higgs doublet field $\Phi$ 
does have the VEV $\langle \Phi^0 \rangle =v/\sqrt{2}$, where $v \simeq 246$ GeV. 
Therefore, only $v$ gives the mass to the quarks and leptons. 

The origin of the dimension five operator in Eq.~(\ref{eq:Leff}) would be in the context of composite Higgs models~\cite{composite}. 
For example, in the $SO(6)/SO(5)$ model where the global symmetry $SO(6)$ is spontaneously broken to 
$SO(5)$ at the composite scale,  there are 5 pseudo Nambu-Goldstone bosons (pNGBs). 
They become components of one Higgs doublet and one neutral real scalar singlet filed in the Higgs sector~\cite{so6}.
This singlet couples to charged fermions as
\begin{align}
{\cal L}_{\rm int} = i m_f \sqrt{\frac{\xi}{1-\xi}} \cot \theta_f~ S {\bar f}_L  f_R~, 
\end{align}
where $m_f$ is the mass of the fermion, and $\xi$ is the compositeness parameter, and 
the mixing angle $\cot \theta_f$ is the $Z_2$ breaking parameter 
which vanishes at $\theta_f = \pi/2$ where the exact $Z_2$ symmetry ($S \to -S$) recovers. 

The existence of charged singlet scalars would also be realized in the context of composite Higgs models.  
In the composite Higgs models, the number of light scalar degrees of freedom in the low energy effective theory is determined by the 
symmetry breaking structure $G/H$.
A part of the list for various $G/H$ and corresponding extended Higgs sectors is presented 
in Ref.~\cite{compositelist}. 
For example, in the $(SO(6))^2/SO(6)$ model, where the global symmetry $(SO(6))^2$ 
is spontaneously broken to $SO(6)$, it appear 15 pNGBs which are decomposed as 
one real singlet, two doublets and two real triplet fields~\cite{so6sq}.  
In such a model, we have the singlet real scalar field $S$ which couples to 
both $t_L \overline{t}_R$ and a number of (singly) charged scalar bosons.

\begin{figure}[t]
\begin{center}
\includegraphics[width=70mm]{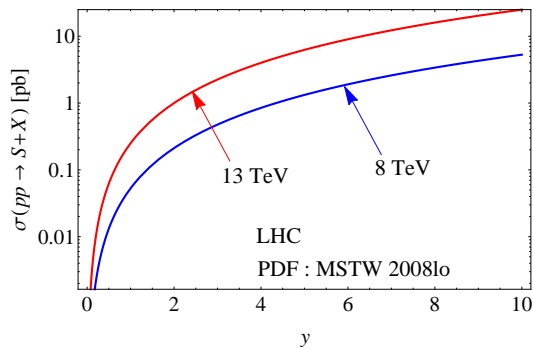}
\end{center}
\caption{The hadronic production cross section $pp \to S +X$ as a function of $y$ at the LHC 
for the centre-of-mass energy to be 13 TeV (red) and 8 TeV (blue). }
\label{fig:ggs}
\end{figure}

In the following, however, we do not specify the fundamental model which predicts the Lagrangian 
in Eq.~(\ref{eq:Leff}). Instead, we consider models with extended scalar sectors with the 
interaction in  Eq.~(\ref{eq:Leff}) in a general framework and try to explain the excess at 750 GeV 
in the recent LHC data. 

For simplicity, we here consider the case where the trilinear scalar couplings $\mu_q^{ij}$ in Eq.~(\ref{eq:Leff}) 
are diagonal and universal, and all  charged scalars are degenerate in mass,  
\begin{align}
 \mu_q^{ij}  &= \mu \delta^{ij},  \\
  m_{\phi_i^{\pm q}}  &= m_{\pm},  \hspace{6mm} \forall q,
 \end{align}
 so that there are only two coupling parameters $y$ and $\mu$ as well as the common mass 
 of charged scalars $m_{\pm}$. 
  
The basic idea of our scenario is the following.  We assume that the excess is the 
result of the production and decay of $S$ with the mass of 750 GeV.  
The production cross section $pp \to S +X$ is dominated by gluon  fusion of top-quark loop mediation.   
The main decay mode of $S$ is $t \overline{t}$, so that the production cross section of 
$\sigma(pp\to S X)$ and the total width $\Gamma_S$ are correlated and controlled by 
the coupling $y$. From the observed value of $\Gamma_S$, the magnitude of $y$ is determined. 
On the other hand, the decay rate of $S \to \gamma\gamma$ is determined 
by the top-loop contribution and also by the charged-scalar loop effect. 
If the number of charged scalars is not small, scalar loop contributions dominate 
the top-loop effect. In such a case, the decay rate is determined by the trilinear scalar coupling $\mu$. 
In the following, we show that the data of the excess can be explained 
by tuning these parameters under the constraint from the 8 TeV data.
 
\section{Numerical Evaluation}

The partonic production cross section $gg \to S$  is given by 
\begin{align}
\hat{\sigma}(gg \to S)
= \frac{G_F \alpha_S^2(\sqrt{\hat{s}})}{288 \sqrt{2} \pi}
\left|
\frac{3}{4} F_f(\tau_t) 
\right|^2
\times \frac{y v^2}{\sqrt{2} m_f \Lambda }~,
\end{align}
where $\hat{s}$ is the centre-of-mass energy of this subprocess 
and $\tau_t = 4 m_t^2/\hat{s}$. 
The hadronic cross section is evaluated by 
\begin{align}
\sigma (pp \to S  X) = K
\int^1_{\tau_S = m_S^2/s} d \tau \frac{d {\cal L}_{gg}}{d \tau}\times \hat{\sigma}(gg \to S)~,
\end{align}
where $d {\cal L}_{gg}/d \tau $ is the luminosity function. 
We here use MSTW2008 LO~\cite{Martin:2009iq}, and the $K$ factor is taken to be 2.5 in our calculation~\cite{k-factor}. 
When we take $\Lambda = v$, we obtain 
\begin{align}
\sigma(pp \to S) \simeq 0.3  y^2~{\rm pb}~.
\end{align}
for $\sqrt{s}=13$ TeV.  
In Fig.~\ref{fig:ggs}, the production cross section of $pp \to S+X$ is shown at the leading order 
as a function of $y$ for $\sqrt{s}=13$ TeV (red) and $8$ TeV (blue). 

We next consider the decay branching ratios of the produced $S$. 
We here assume that $m_S < 2 m_\pm$ so that 
the charged scalars affect the total width $\Gamma_S$ only via the 
quantum loop contributions in $S\to \gamma\gamma$.  
The decay rates of the singlet $S$ are calculated by 
\begin{align}
\Gamma(S \to t \bar{t})
& = 
 \frac{N_c g^2 m_t^2}{32 \pi m_W^2} m_S^{} |\kappa_{stt}|^2
\left(
  1 - \frac{4 m_t^2}{m_{S}^2}
\right)^{\frac{3}{2}},  \\
\Gamma (S \to \gamma \gamma ) 
& = 
\frac{\alpha^2 g^2}{1024 \pi^3} \frac{m_S^3}{m_W^2} \nonumber \\
&
\hspace*{-0.8cm} \times \left| 
\kappa_{stt} \frac{4}{3} F_t \left( \frac{4 m_t^2}{m_S^2} \right)
 + r \kappa_{s \pm}  F_0 \left( \frac{4 m_{\pm}^2}{m_S^2} \right)
\right|^2~,  \\
\Gamma(S\to Z\gamma)
&=\frac{\alpha^2m_S^3}{128\pi^3v^2}
\left(1-\frac{m_Z^2}{m_S^2}\right)^3\nonumber\\ 
& \hspace{-0.8cm} \times \left|
2\kappa_{stt}J_f
+\sum_i \frac{q_ig}{c_W}\left(I_3^i-s_W^2q_i\right)
\frac{\mu}{v}J_S
\right|^2, 
\end{align}
where $r = \sum_{q} q^2 n_q$, and 
\begin{align}
\kappa_{stt}   & = \frac{y v}{\sqrt{2} m_t}~,~~~~
\kappa_{s \pm} = \frac{m_W}{g m_{\pm}^2} \mu~.
\end{align}

\begin{widetext}

\begin{figure}[t]
  \begin{center}
    \begin{tabular}{ccc}
 \resizebox{50mm}{!}{\includegraphics{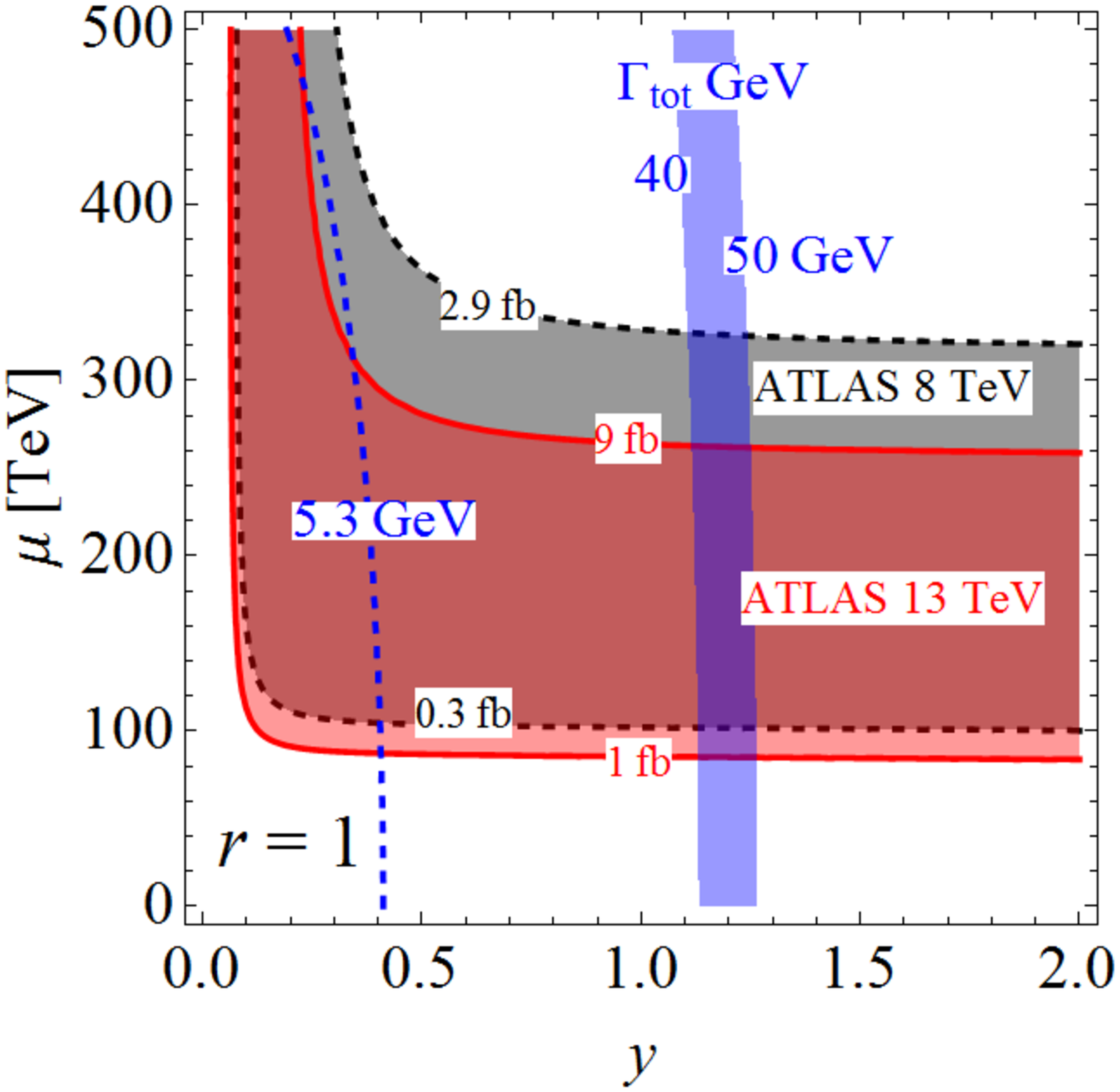}} &
      \resizebox{50mm}{!}{\includegraphics{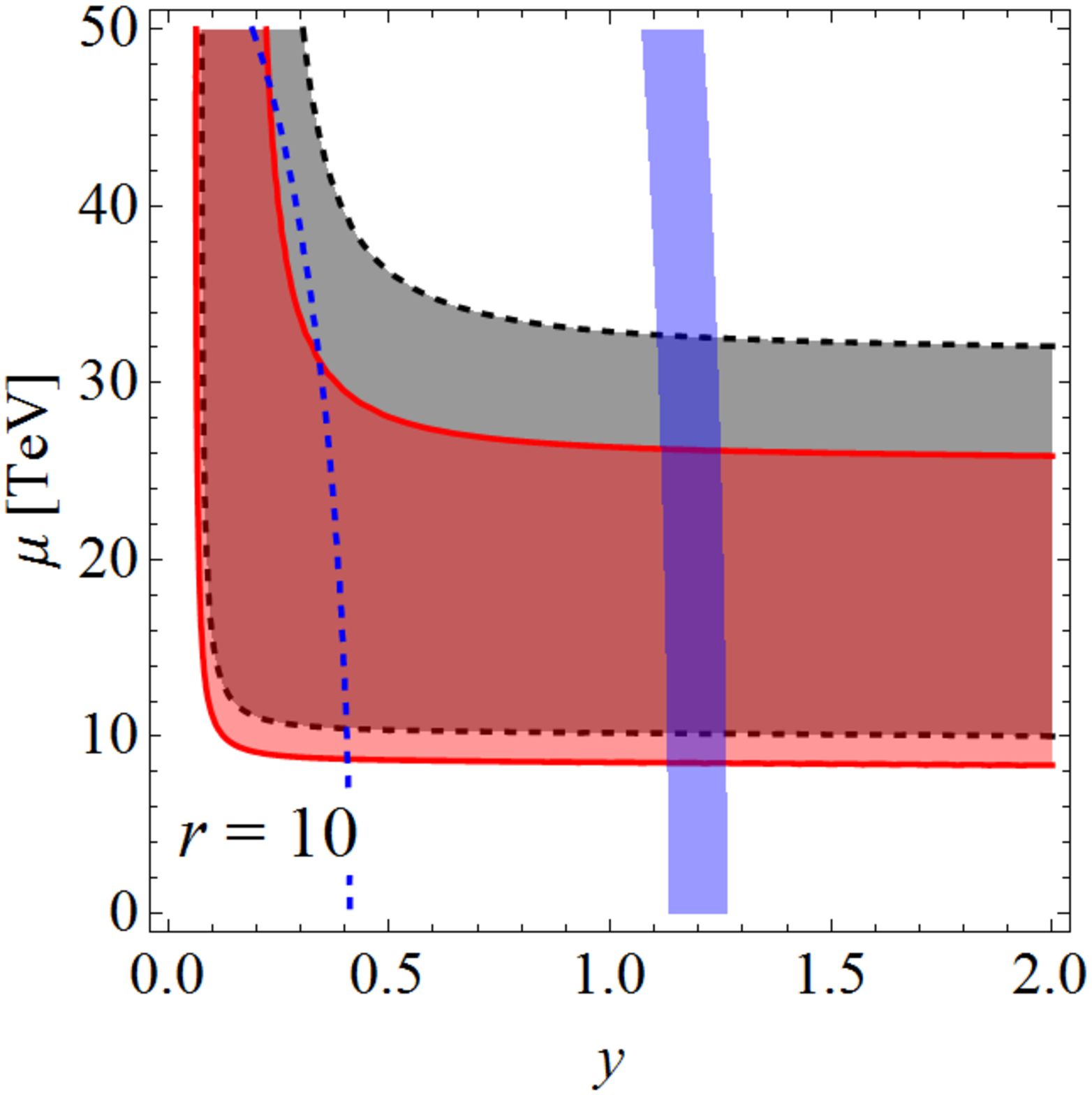}} & 
      \resizebox{50mm}{!}{\includegraphics{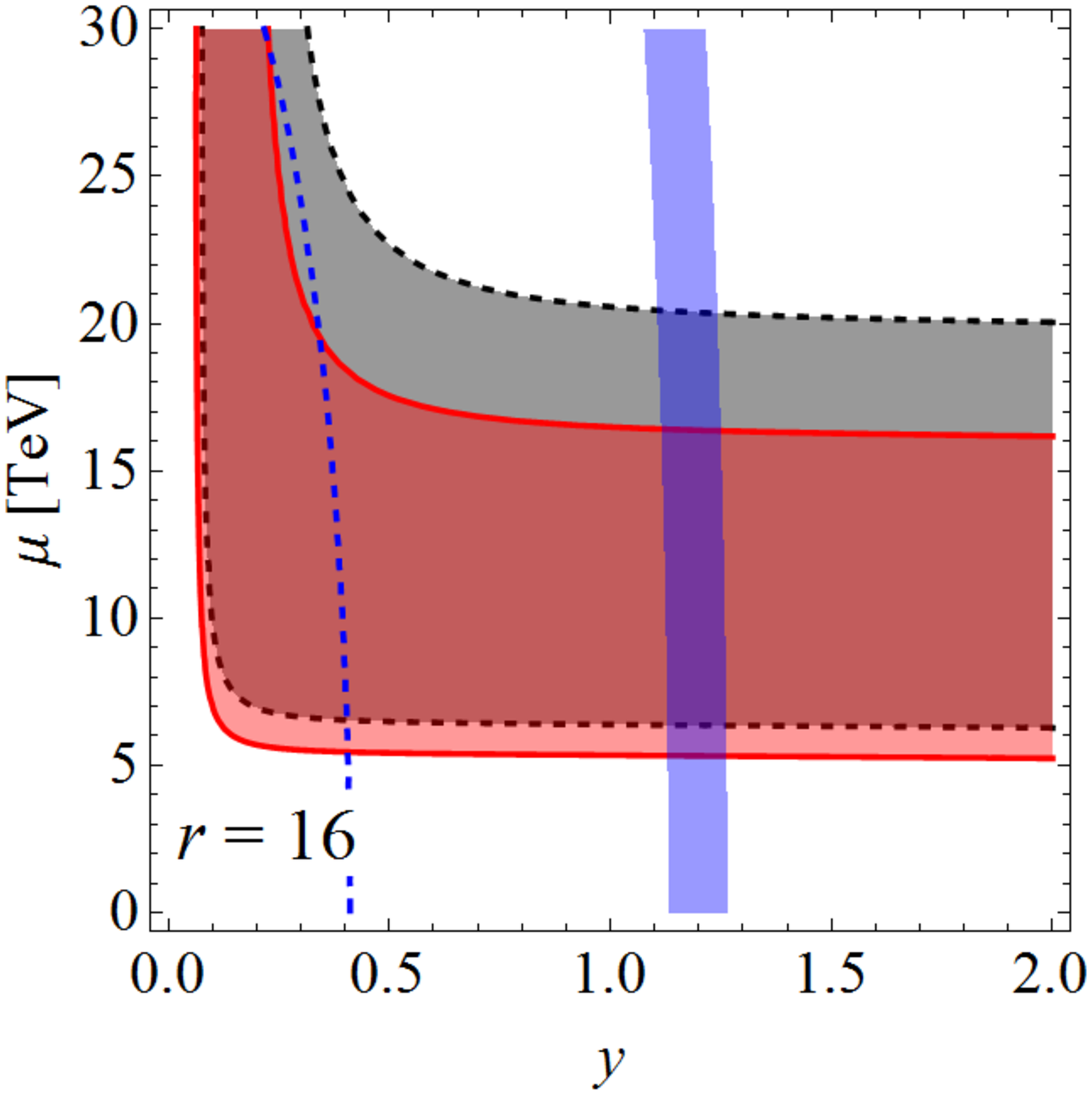}} \\
      \resizebox{50mm}{!}{\includegraphics{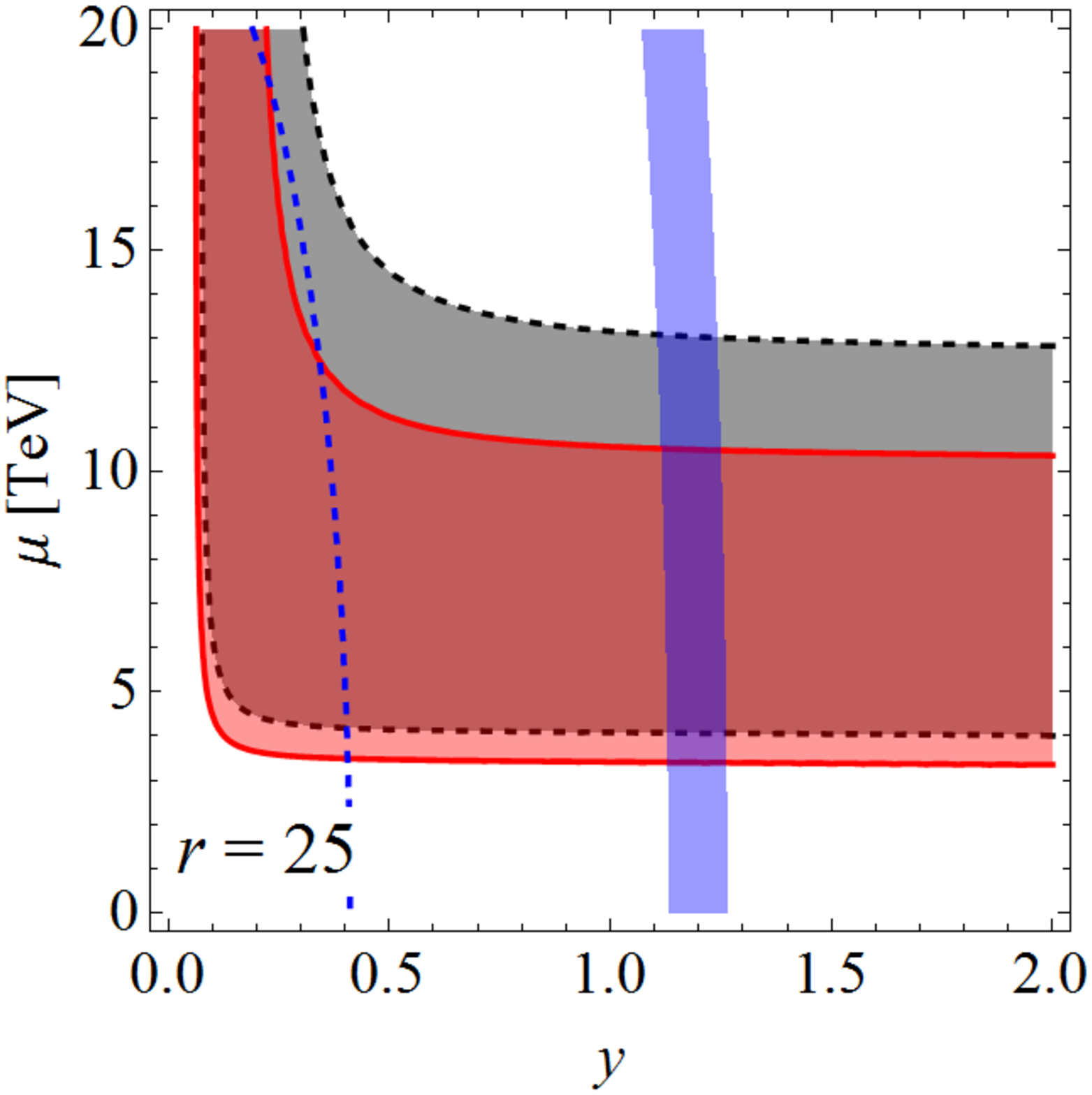}}  &
      \resizebox{50mm}{!}{\includegraphics{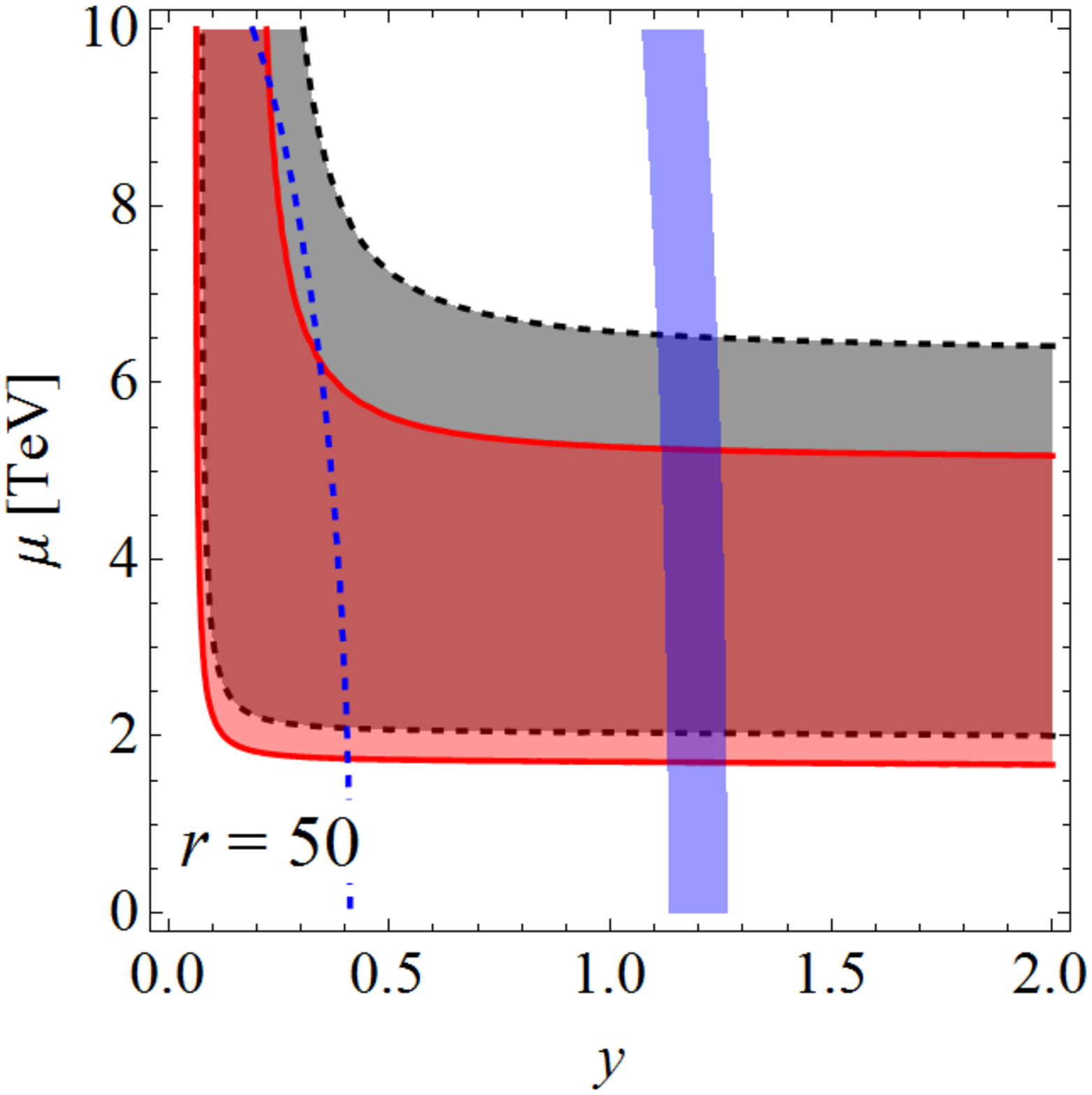}} &         
      \resizebox{50mm}{!}{\includegraphics{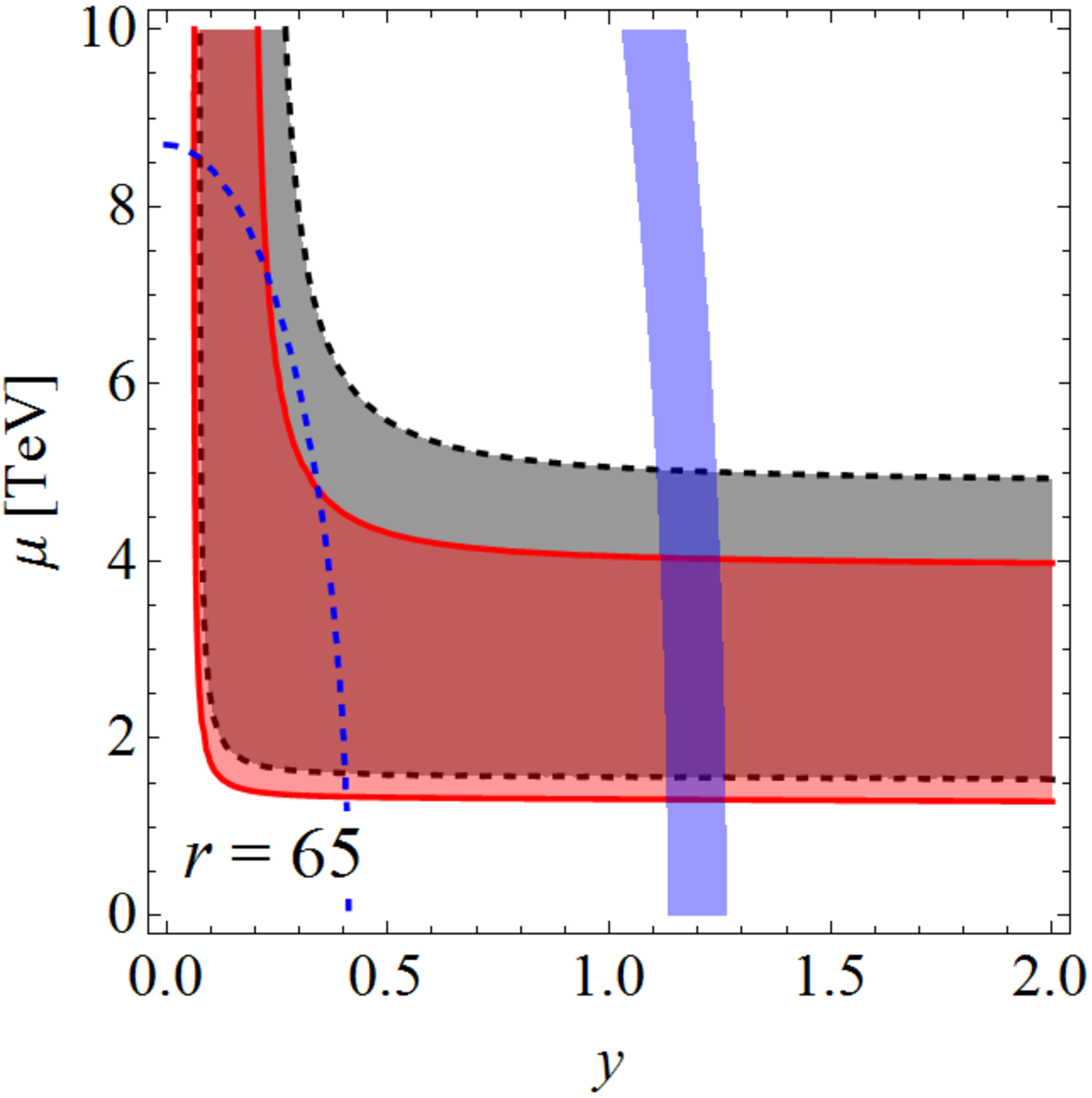}} \\
    \end{tabular}
    \caption{Contour plots of the signal cross section $\sigma(pp \to S X)\times {\rm Br}(S \to \gamma \gamma)$ [fb] on the $y$-$\mu$ plane 
     for $m_\pm =400$ GeV.
    The regions which satisfy the 13 TeV data (red)~\cite{ATLAS2} and 8 TeV data (grey)~\cite{ATLAS1} within the 95 \% CL are shown.  
    The regions  where the total width of $S$ is in 40 GeV $< \Gamma_S < 50$ GeV (blue band) and the curve of 
     $\Gamma_S = 5.3$ GeV (blue dashed), the energy resolution of the diphoton system,  are also indicated.   
     From top-left to bottom-right, the results are shown in the models wish $r=1$, $10$, $16$, $25$, $50$ and $65$. }
         \label{fig:allowed_400}
  \end{center}
\end{figure}

\end{widetext}

\noindent
The loop integral functions are defined by~\cite{integral} 
\begin{align}
&F_t(\tau) =
  - 2 \tau 
\left[
1 + (1-\tau)f(\tau)
\right], \\
&F_0 (\tau)= 
\tau (1 - \tau f(\tau)),  
\end{align}
and 
\begin{align}
f(\tau)
=  
\left\{
\begin{array}{l}
\left[ \sin^{-1} \frac{1}{\sqrt{\tau}} \right]^2~~~{\rm for}~\tau \geq 1, \\
- \frac{1}{4} 
\left[
\log \left\{ \frac{1 + \sqrt{1-\tau}}{1 - \sqrt{1-\tau}} \right\} - i \pi
\right]^2,~~~{\rm for}~\tau < 1. 
\\
\end{array}
\right. 
\end{align}
$J_f$ is written by $A_{1/2}^H$ in Ref.~\cite{Djouadi:2005gi} as
\begin{equation}
J_f=\frac{v_f}{s_Wc_W}A_{1/2}^H\;.
\end{equation}
The loop contribution $J_S$ is given in 
Ref.~\cite{JSdefine}.
Our model can be classified by the parameter $r$,  and 
characterized by the parameters of $y$, $\mu$ and $m_{\pm}$ with $m_S$ being set to be 750 GeV.  

The signal cross section of $pp \to S + X \to \gamma \gamma + X$ is then given by
\begin{align}
& \sigma (pp \to S X \to \gamma \gamma X)
= \nonumber \\ 
& \hspace*{1cm}\sigma(pp \to S X) \cdot {\rm Br}(S \to \gamma \gamma), 
\end{align}
where we employ the narrow width approximation, because the 
ratio $\Gamma_S/m_S$ is smaller than 0.1. 

Now we survey the parameter region where the data of the excess at 750 GeV is 
explained without contradiction with the Run I data for the process 
 $pp \to S + X \to \gamma \gamma + X$ at 8 TeV~\cite{ATLAS1, CMS1};   
\begin{align}
{\rm ATLAS}~:~
& \sigma(pp \to S X \to \gamma \gamma X)
  \simeq 1.6 \pm 1.3~{\rm fb}~(95\% {\rm CL})~,\nonumber \\
{\rm CMS}  ~:~
& \sigma(pp \to S X \to \gamma \gamma X)
\simeq 0.9 \pm 0.6~{\rm fb}~(95\% {\rm CL})~. 
\end{align}
If the model contains doubly charged scalar bosons, 
we have to take into account the constraint on the mass from the LHC data. 
In particular, if the doubly charged scalars from isospin singlets (triplets)
decay into dilepton, the current lower bound is about 430 GeV (550 GeV)~\cite{dilepton}.  
On the other hand, if they are of the complex triplet scalar fields, 
it can mainly decay into diboson ($W^\pm W^\pm$) when the VEV for the triplet is larger than 0.1 MeV.  
In 

\begin{widetext}

\begin{figure}[t]
  \begin{center}
    \begin{tabular}{ccc}
 \resizebox{50mm}{!}{\includegraphics{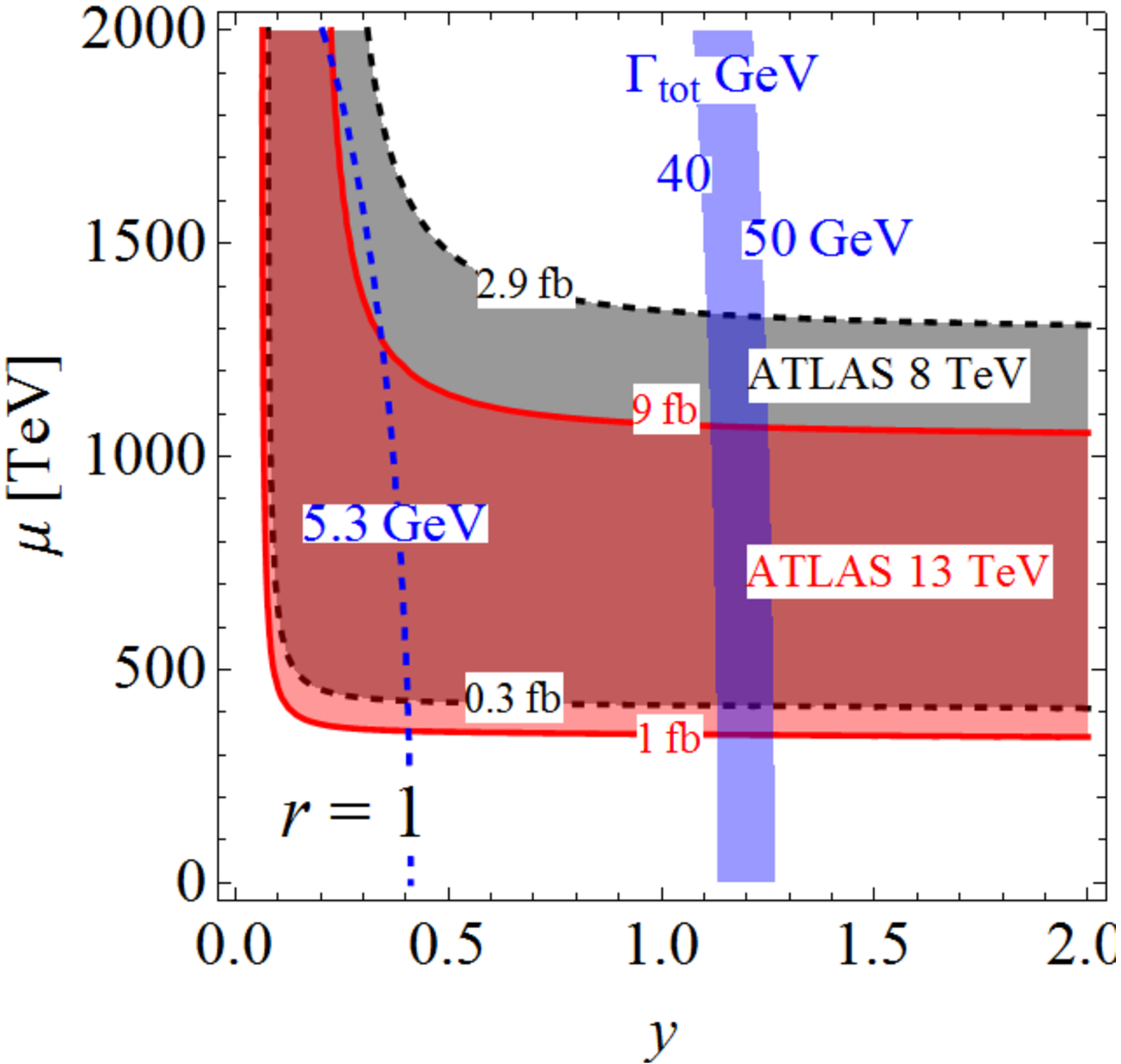}} &
      \resizebox{50mm}{!}{\includegraphics{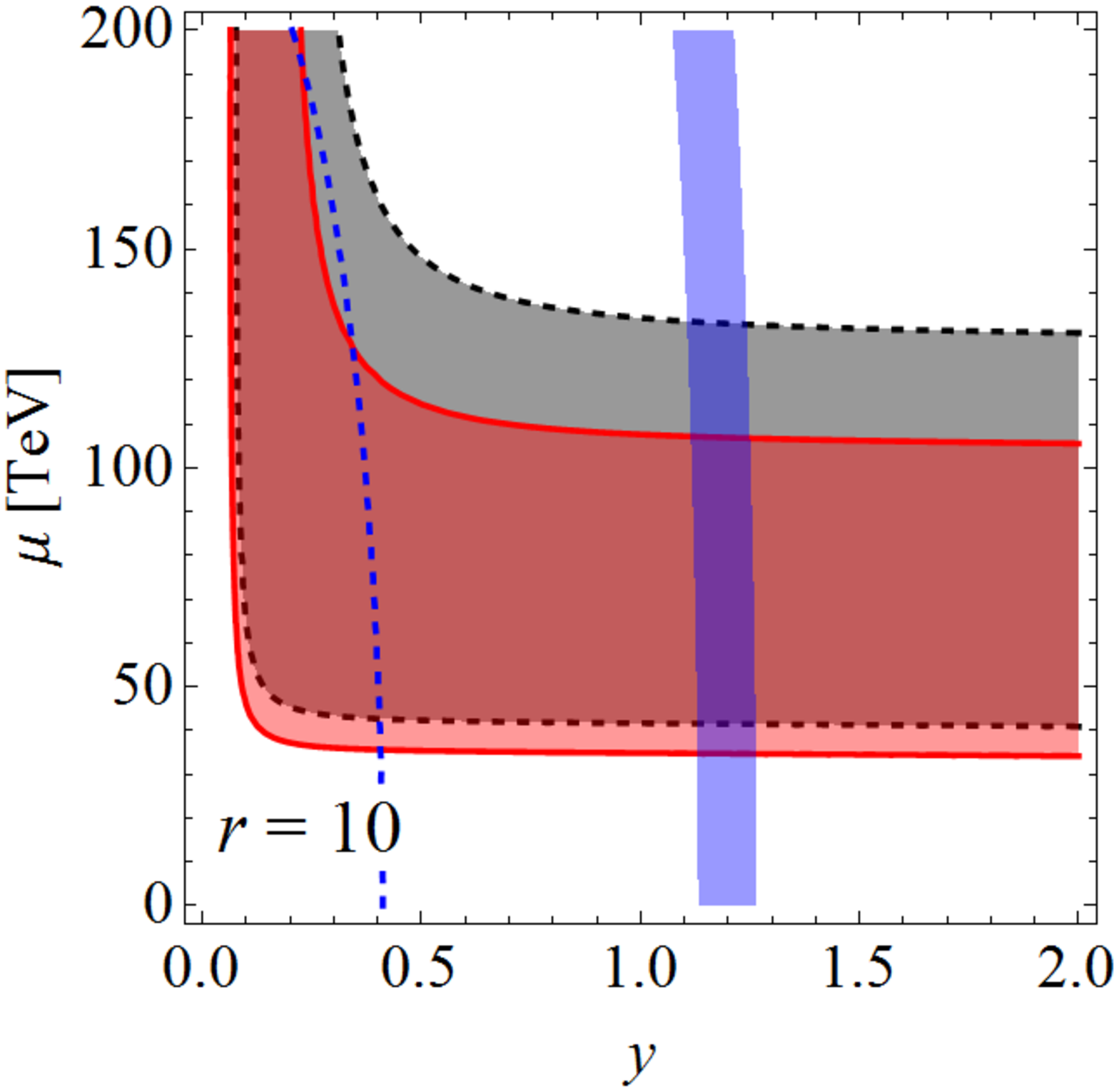}} & 
      \resizebox{50mm}{!}{\includegraphics{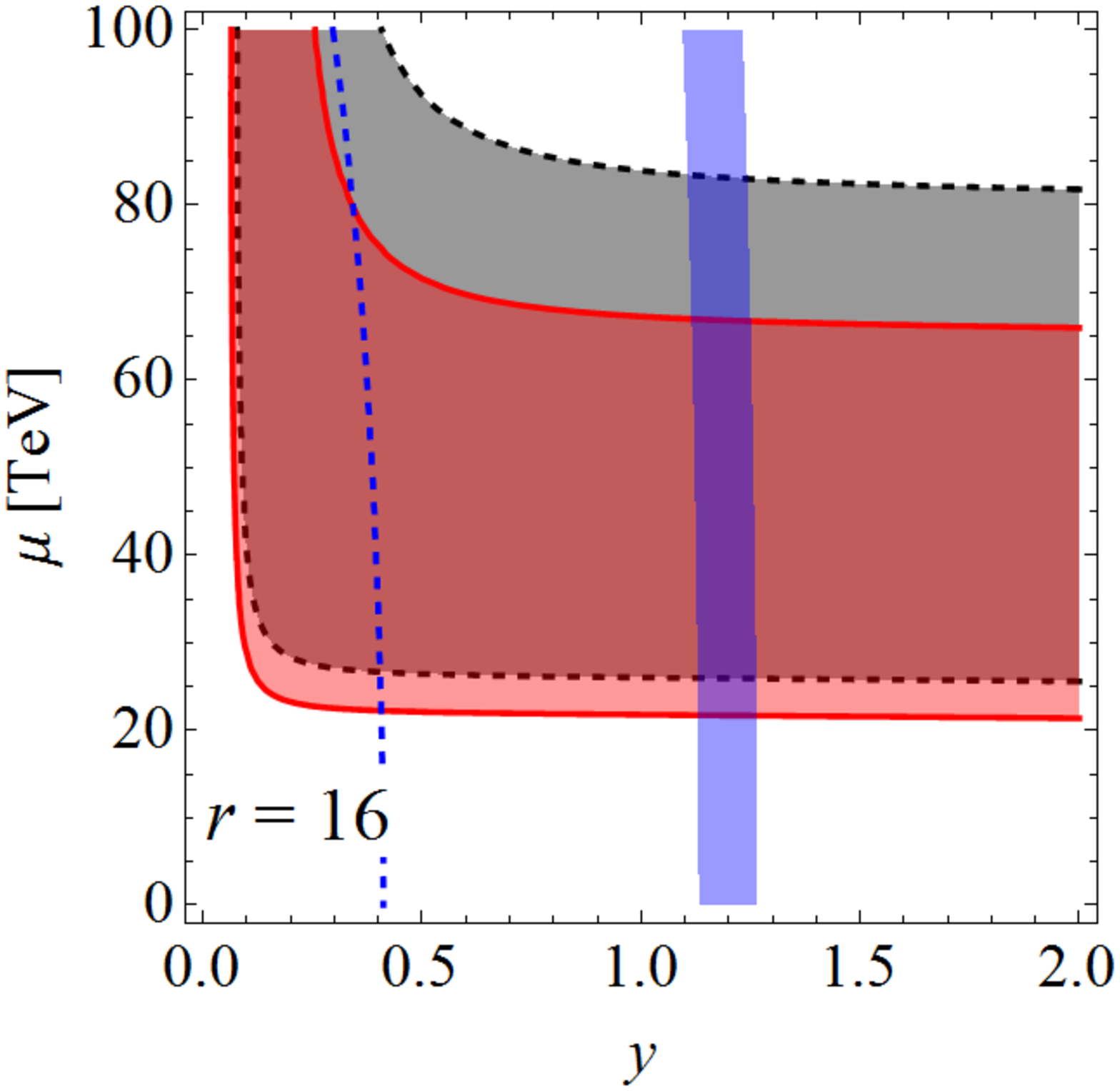}} \\
      \resizebox{50mm}{!}{\includegraphics{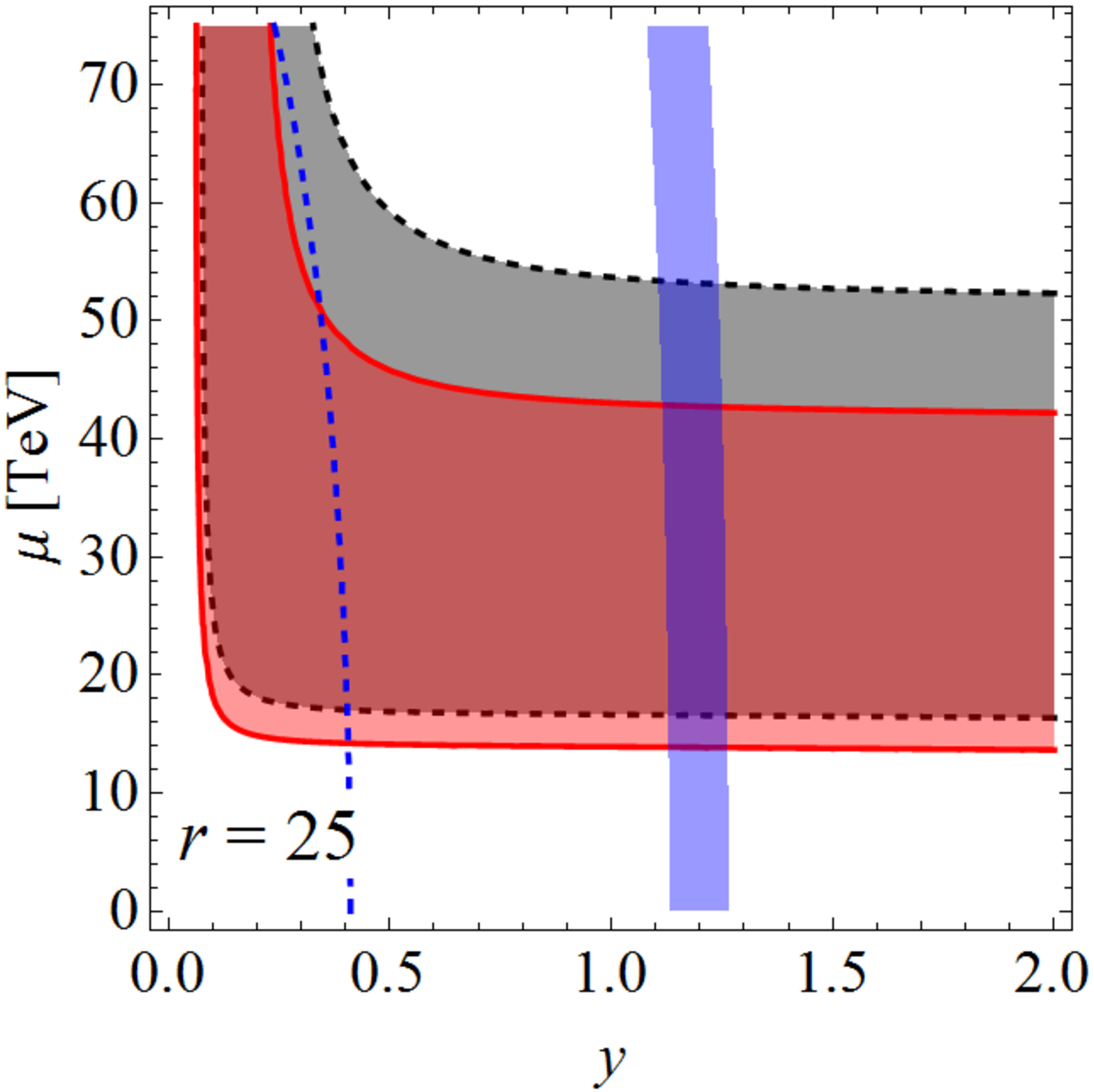}}  &
      \resizebox{50mm}{!}{\includegraphics{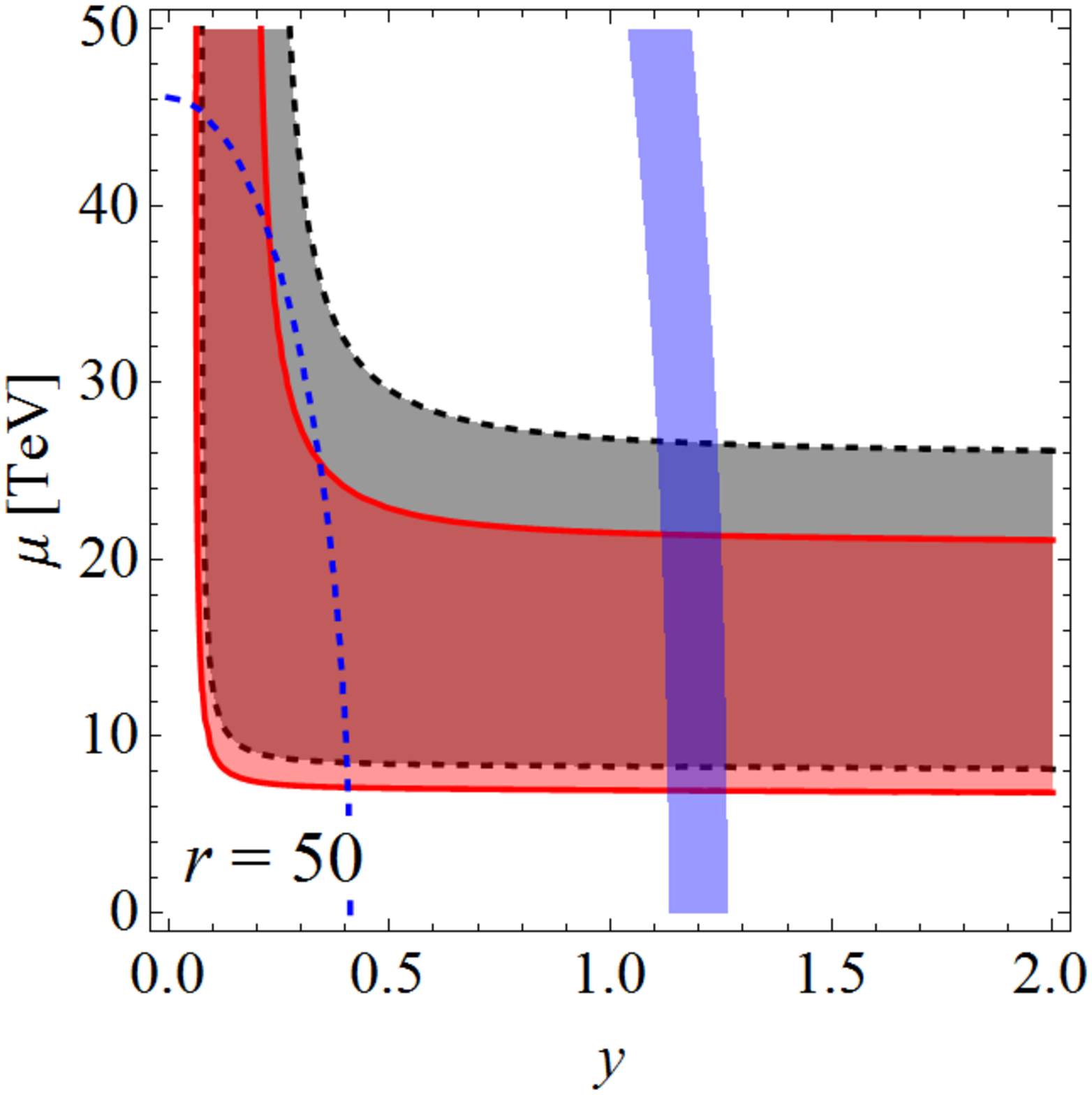}} &         
      \resizebox{50mm}{!}{\includegraphics{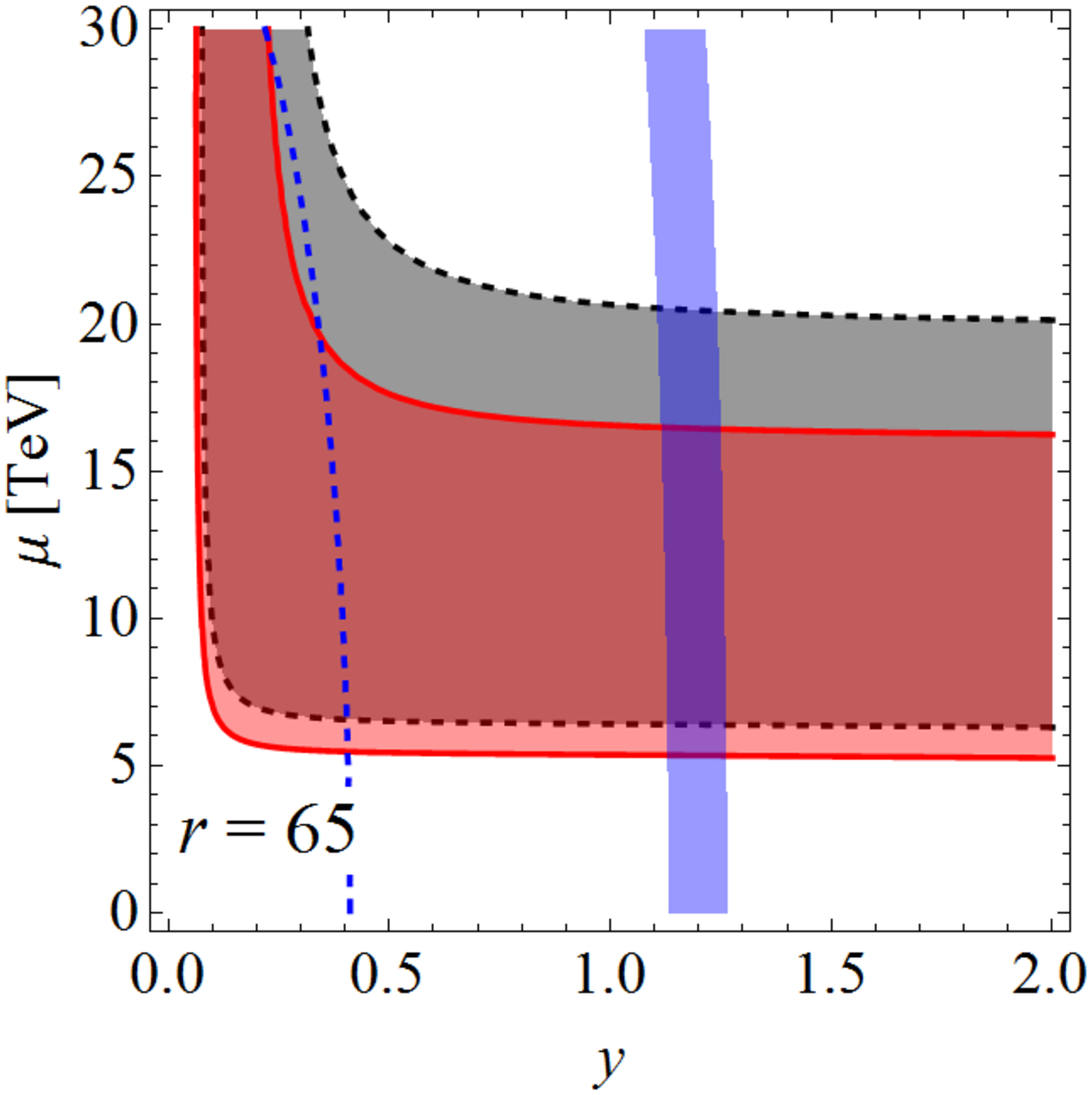}} \\
    \end{tabular}
    \caption{Contour plots of the signal cross section $\sigma(pp \to S X)\times {\rm Br}(S \to \gamma \gamma)$ [fb] on the $y$-$\mu$ plane 
     for $m_\pm =600$ GeV.
    The regions which satisfy the 13 TeV data (red)~\cite{ATLAS2} and 8 TeV data (grey)~\cite{ATLAS1} within the 95 \% CL are shown.  
    The regions  where the total width of $S$ is in 40 GeV $< \Gamma_S < 50$ GeV (blue band) and the curve of 
     $\Gamma_S = 5.3$ GeV (blue dashed), the energy resolution of the diphoton system,  are also indicated.   
     From top-left to bottom-right, the results are shown in the models wish $r=1$, $10$, $16$, $25$, $50$ and $65$.}
    \label{fig:allowed_600}
  \end{center}
\end{figure}

\end{widetext}

\noindent
such a case, 
the current mass lower limit is about 90 GeV~\cite{diboson}, 
which is much relaxed as compared to the case of dilepton decays. 
If we consider $n_q$ ($> 1$) of doubly charged Higgs bosons, these bounds should become 
stronger.  We here do not specify the isospin of charged scalar bosons and also their main decay mode for a while, 
and we come back to this issue later.

In Fig.~\ref{fig:allowed_400}, 
contour plots of the regions satisfying the data from the 13 TeV Run (red)~\cite{ATLAS2}  and those at 8 TeV (grey)~\cite{ATLAS1} are 
shown (in the 95 \% CL) on the $y$-$\mu$ plane for the six models with $r=1$, $10$, $16$ $25$, $50$, and $65$. 
The region where the width of $S$ is 40 GeV $ < \Gamma_S < 50$ GeV is indicated 
by blue shaded regions.  We also draw the curve of $\Gamma_S=5.3$ GeV, the current 
resolution for the diphoton system~\cite{ATLAS2}. 
The universal mass of charged scalars is set to be 400 GeV.  
In each model,  there is the region where all data are satisfied. 
For smaller $r$, relatively large $\mu$ is required to satisfy the data, while 
for relatively large number of $r$, $\mu$ can be lowered to a few TeV. 
For instance, for the model with $r=1$ where only one singly charged scalar filed 
is introduced, the required value of $\mu$ is 100-200 TeV to satisfy the data. 
On the other hand,  for the model with $r=65$ which corresponds to the models 
with $(n_1, n_2) = (65, 0)$, $(45, 5)$, $(25, 10)$, $(13, 13)$ etc, 
the required value of $\mu$  is at most a few TeV.

In Fig.~\ref{fig:allowed_600}, the similar figures for the results 
with the mass of the universal charged scalar mass $m_\pm$ to be 600 GeV are shown 
with the same fashion. 
We see that the required values of $\mu$ for each model are larger than the cases with 
$m_\pm =400$ GeV.

\section{Discussion}

We here discuss theoretical constraints which limit parameter regions, and then   
give some comments on the relation of our model to the new physics phenomena.  

First, in order to obtain enough enhancement in the diphoton decay rate of the singlet $S$, 
a larger value of $\mu$ is required for a smaller values of $r$. 
However, taking a too large value of $\mu$ compared to $m_S$ and $m_{\pm}$ 
possibly causes dangerous charge breaking minima.
For the case of $m_S=750$~GeV and $m_{\pm}=400\text{--}600$~GeV, the value of 
$\mu$ larger than about 10~TeV is not favored at all  by the similar analysis to the case of large 
trilinear coupling in the minimal supersymmetric standard model (MSSM)\cite{CCBinMSSM}.

Second, the perturbative unitarity bound for scattering processes such as 
$\phi^{+q}\phi^{-q}\to \phi^{+q}\phi^{-q}$ should also be taken into account in the case with a large $\mu$. 
It is known that in the MSSM, constraints from perturbative unitarity for the trilinear coupling 
are similar in strength to bounds from color and charge breaking minima\cite{UBinMSSM}.
The unitarity bound on $\mu$ in our model is also expected to lead to a similar
constraint from charge non-breaking vacuum.   

Therefore, from these theoretical constraints, the models with a small $r$ and a large $m_\pm$ are 
not favored even though there are regions which satisfy the data of the excess. 
As seen in Fig.~\ref{fig:allowed_400}, for $m_{\pm} =400$ GeV the cases with $r=25$, $50$ and $65$ can be safe 
from these theoretical bounds, while  for $m_{\pm} =600$ GeV only those with  $r=65$ can be allowed 
(see Fig.~\ref{fig:allowed_600}). 

In order to have a relatively large values of $r$, introduction of a scalar field with a higher isospin representation would be 
helpful.  For example, in the model with an isospin septet scalar field, there are many multiply charged 
scalar bosons $\phi^{\pm 5}$, $\phi^{\pm 4}$, $\phi^{\pm 3}$, $\phi^{\pm 2}$, $\phi^{\pm}$ and $\overline{\phi}^\pm$, 
which give $r=56$.  The phenomenology of the septet field is discussed in Ref.~\cite{septet}.  
We note that models with a higher representation scalar field than the septet are not realistic from viewpoint 
of perturbative unitarity~\cite{logan}. 

In our analysis, we only have considered the models with only one real singlet field $S$. However, it would also be possible 
to consider the cases with more real singlets $S_i$ ($i=1, \cdots N$).  
If they have the common mass and the universal couplings with $t_L \overline{t}_R$ and with charged scalars, then 
the signal cross section becomes $N^2$ larger than the case with $N=1$. 
In such a case, the magnitude of the trilinear coupling can be smaller so that the constraint from 
perturbative unitarity and charge non-breaking would be milder. 
In this case, the excess at 750 GeV can be explained with smaller values of $r$.

In our scenario, many (multiply) charged scalar fields are introduced. 
Such introduction of many scalars can also be seen in the models for quantum generation of 
tiny neutrino masses (so called radiative seesaw scenarios), where 
neutrino masses are deduced from the extended scalar sectors at one-loop~\cite{zee,ma}, two-loop~\cite{zeebabu} 
and three-loop levels~\cite{knt, aks, gnr}.  
Therefore, the excess at 750 GeV would be indirect evidence for such radiative seesaw scenarios with $S$ which couples to $t_L \overline{t}_R$. 
In addition, introduction of many scalars can cause strongly first order phase transition at electroweak symmetry breaking~\cite{1stOPT}, which 
is required for a successful scenario of electroweak baryogenesis~\cite{ewbg}. 

For $m_S =750$ GeV and the observed value of $\Gamma_S \sim 45$ GeV, the branching ratio of $S \to \gamma\gamma$ is required to take values  
between about 0.6 \% and about 3.5 \% to satisfy all the diphoton data for all cases of $r$ and $m_\pm$.    
The data can be satisfied with similar values of the branching ratio even if the value of the width is smaller than 45 GeV. 
The main decay mode of $S$ is always $tt$, whose branching ratio is larger than 95 \%. 
Produced number of $t \overline{t}$ via the $S$ decay is much smaller 
than the uncertainty  in the data of the $t \overline{t}$ production cross section at the 8 TeV~\cite{tt_1} and the 13 TeV~\cite{tt_2}. 

In the following, we mention some phenomenological features with speculation. Detailed study 
is beyond the scope of this letter. 
If our scenario is true, the second phenomenological signature would be 
the discovery of charged Higgs bosons with the mass to be 400-600 GeV.  
In particular, if some of them have double electric charge, the final state would be dilepton or diboson, 
depending on their isospin charges and the other parameters.  
If they decay into dilepton, the signature is expected to be observed very soon at LHC Run II, 
otherwise the model is ruled out.   
If doubly charged scalars have isospin, the main decay mode can be same sign diboson. 
In such a case, the current lower limit is about 90 GeV~\cite{diboson}. In order to detect the signal 
around 400-600 GeV, considerable amount of luminosity has to be accumulated. 
Finally if the model contains higher order isospin representation scalar fields, their charged scalar components 
also enhance the $Z\gamma$ decay rate by the loop effect, whose branching ratio can be 
a few times 1 \%.  In such a case, the signal of $S\to Z\gamma$ would be discovered in the near future at LHC Run II.

\section{Conclusion}

We have discussed extended scalar models which can explain 
the recent results of diphoton excess at 750 GeV at LHC Run II.      
An additional singlet scalar boson with the mass of 750 GeV, which couples to top quarks via dimension five 
operator, is produced via gluon fusion and decays into two photons via loop 
contributions of a number of (multiply) charged scalar bosons. 
The excess can be explained without contradicting the data from LHC Run I. 
From theoretical consistencies such as perturbative unitarity and charge non-breaking, 
the number of charged particles can be constrained. \\

\noindent 
{\it Acknowledgments}: \\
We would like to thank Kunio Kaneta for useful discussions. 
This work was supported by 
Grant-in-Aid for Scientific Research from the MEXT, Japan, Nos.\ 23104006 (S. K.) and 23104011 (T. S.), 
and Grant H2020-MSCA-RISE-2014~no.~645722 (Non Minimal Higgs) (S. K.).

\end{document}